\documentclass{article}

\input{tcilatex}

\begin{document}

\begin{center}
\textbf{ENERGETICS IN\ CONDENSATE\ STAR\ AND WORMHOLES}

\bigskip

K.K. Nandi$^{1,2,3,a}$, Y.Z. Zhang$^{2,b}$, R.G. Cai$^{2,c}$ and A. Panchenko%
$^{3,d}$

$^{1}$Department of Mathematics, University of North Bengal, Siliguri (WB)
734013, India

$^{2}$Institute of Theoretical Physics, Chinese Academy of Sciences, Post
Box 2735, Beijing 100080, China

$^{3}$Joint Research Laboratory, Bashkir State Pedagogical University, Ufa
450000, Russia

$\bigskip $

$^{a}$Email: kamalnandi1952@yahoo.co.in

$^{b}$Email: zyz@itp.ac.cn

$^{c}$Email: cairg@itp.ac.cn

$^{d}$Email: alex.souljaa@gmail.com

\bigskip

\textbf{Abstract}
\end{center}

It is known that the total gravitational energy in localized sources having
static spherical symmetry and satisfying energy conditions is negative
(attractive gravity). A natural query is how the gravitational energy
behaves under circumstances where energy conditions are violated. To answer
this, the known expression for the gravitational energy is suitably adapted
to account for situations like the ones occurring in wormhole spacetime. It
is then exemplified that in many cases the modified expression\ yields
desirable answers. The implications are discussed.

\bigskip

PACS\ numbers: 04.20-q, 04.20.Cv, 95.30.Sf

\begin{center}
\bigskip \textbf{1. Introduction}
\end{center}

\bigskip Classical wormholes, just as black holes, represent self consistent
solutions of Einstein's theory of general relativity. Topologically they are
like handles connecting two distant regions of spacetime. Wormhole solutions
were conceived as particle models by Einstein himself (Einstein-Rosen bridge
[1]) in 1935. (A 1916 predecessor of wormholes is Flamm [2] paraboloid). The
seminal theoretical framework laid in 1988 by Morris, Thorne and Yurtsever
(MTY) [3] has since led to serious investigations into the topic of wormhole
physics. In addition to the traditional method of solving Einstein's
equations, there exist what is well known as Synge's method (It is a reverse
method employed by MTY in constructing wormholes) in which one first fixes
the spacetime geometry and then computes, via field equations, the stress
components needed to support such a geometry. The resulting stress
components automatically satisfy local conservation laws in virtue of
Bianchi identities. Either method has led to several wormhole solutions in
well known theories such as in Brans-Dicke theory [4], scalar field theory
with potential [5], low energy string theory [6], braneworld model [7],
phantom model [8], Chaplygin gas model [9], Thin-Shell model [10] and in
cosmology [11]. Configurations resulting from these theories could be
potential candidates to occur in a natural way and are (perhaps) of some
astrophysical interest [12,13]. A largely unnoticed but important work was
carried out in 1948 by Fisher [14] who discovered formal solutions to
minimally coupled scalar field Einstein equations with a positive sign
kinetic term. Thereafter, in 1973, Ellis [15] and independently, Bronnikov
[16] found wormhole solutions of the Einstein minimally coupled theory with
a negative sign kinetic term. All wormhole solutions require exotic material
for their construction. However, to our knowledge the gravitational energy
content in the interior of exotic matter distribution has not yet been
studied. An initiative along this direction can be taken by employing the
formulation of gravitational energy provided by the works of Lynden-Bell,
Katz and Bi\v{c}\'{a}k [17-19]. A Maxwellian analogy together with a
conformal factor interpretation of gravitational energy density, which is
new, is also given in [17].

The energy formulation in references [17-19] is intended for isolating and
calculating the total attractive gravitational energy $E_{G}$ of stationary
gravity fields. In our view, their formulation did not require any
compelling restriction on the energy conditions of the source matter. We
shall therefore allow the source matter to violate one or more energy
conditions, particularly the Weak Energy Condition (WEC) $\rho >0$ and/or
the Null Energy Condition (NEC) $\rho +p_{r}\geq 0$ where $\rho $ is the
matter energy density and $p_{r}$ is the radial pressure. (Transverse
pressures $p_{\perp }$ are not considered as they refer strictly to ordinary
matter.) The violation of Null Energy Condition (NEC) is a minimal
requirement to have defocussing of light trajectories (\textit{repulsive}
gravity) passing across the wormhole throat [20]. The necessity of NEC
violation in wormholes is provided by the Topological Censorship Theorem
[21] and by dynamical circumstances [22].

In this paper, we first adapt $E_{G}$ to wormhole geometry and distinguish
it by $\widetilde{E}_{G}$. Then we investigate the behavior of $\widetilde{E}%
_{G}$ in certain static spherically symmetric model solutions that violate
the energy conditions as stated above. The examples we consider, one star
and three wormholes, are somewhat critical in nature. Explicit calculations
in the Ellis III wormhole show that the definition of $\widetilde{E}_{G}$ is
robust enough. In the case of de Sitter star, exact calculation shows it has
attractive gravity which then admits a plausible physical interpretation. In
the case of phantom wormhole, we find that $\widetilde{E}_{G}$ is repulsive
around the throat, which is necessary for defocussing effect. Such a
behavior of $\widetilde{E}_{G}$ may serve as a constraint with regard to the
practical feasibility of localized wormholes.

The paper is organized as follows: In Sec.2, we adapt the equation for total
gravitational energy so as to account for the wormhole geometry. In Sec.3,
we work out a specific solution to show that the conformal factor
interpretation holds also in the case of wormhole spacetime. The same
example is used to show the robustness of $\widetilde{E}_{G}$ in Sec.4. In
Sec.5, we apply the energetics to the Mazur-Mottola gravastar [23] and
interpret the result. In Sec.6, we calculate contribution to $\widetilde{E}%
_{G}$ coming from the thin shell. In Sections 7 and 8, we investigate
wormholes which are localized by spacetime cut-off at some radii. Sec.9
summarizes the results. Units are chosen so that $8\pi G=c=1$, unless
specifically restored.

\begin{center}
\textbf{2. Gravitational energy }
\end{center}

We shall consider spherically symmetric static spacetime with the metric
expressed in \textquotedblleft standard" coordinates ($%
x^{0},x^{1},x^{2},x^{3}$) $\equiv $($t,r,\theta ,\psi $) as 
\begin{equation}
ds^{2}=-e^{2\Phi (r)}dt^{2}+g_{rr}(r)dr^{2}+r^{2}(d\theta ^{2}+\sin
^{2}\theta d\psi ^{2}).
\end{equation}%
The stress components resulting from the metric via Einstein's equations are
denoted by 
\begin{equation}
T_{0}^{0}=\rho ,T_{1}^{1}=p_{r},T_{2}^{2}=p_{\theta },T_{3}^{3}=p_{\psi }
\end{equation}%
in which $\rho $ is the matter energy density and $p_{r}$ is the radial
pressure and $p_{\theta }$, $p_{\psi }$ are transverse pressures of the
fluid in its rest frame. Because of spherical symmetry, $p_{\theta }=$ $%
p_{\psi }$.The total gravitational energy $E_{G}$ appropriate for ordinary
matter is given in [17,18] as%
\begin{equation}
E_{G}=Mc^{2}-E_{M}=\frac{1}{2}\int_{0}^{r}[1-(g_{rr})^{\frac{1}{2}%
}]T_{0}^{0}r^{2}dr
\end{equation}%
where the total mass-energy within the standard coordinate radius $r$ is
provided by Einstein's equations as

\begin{equation}
Mc^{2}=\frac{1}{2}\int_{0}^{r}T_{0}^{0}r^{2}dr
\end{equation}%
and the sum of other forms of energy like rest energy, kinetic energy,
internal energy etc is defined by 
\begin{equation}
E_{M}=\frac{1}{2}\int_{0}^{r}T_{0}^{0}(g_{rr})^{\frac{1}{2}}r^{2}dr.
\end{equation}%
The factor $\frac{1}{2}$ comes from $\frac{4\pi }{8\pi }$. The $E_{M}$ is
similar to the geometric definition given by Wald [24]. Since $(g_{rr})^{%
\frac{1}{2}}>1$ by definition (proper radial length larger than the
Euclidean length), one immediately deduces the criteria that $E_{G}<0$
(attractive) if $T_{0}^{0}>0$ [25] and that $E_{G}>0$ (repulsive) if $%
T_{0}^{0}<0.$

For wormhole spacetime, we consider the spherically symmetric spacetime
metric in the generic MTY\ form%
\begin{equation}
ds^{2}=-e^{2\Phi (r)}dt^{2}+\frac{dr^{2}}{1-\frac{b(r)}{r}}+r^{2}(d\theta
^{2}+\sin ^{2}\theta d\psi ^{2})
\end{equation}%
where $\Phi (r)$ and $b(r)$ are redshift and shape functions respectively.
Throughout the wormhole $0<\frac{b(r)}{r}<1$ and that $\frac{b(r)}{r}%
\rightarrow 0$ as $r\rightarrow \infty $. The expressions for the stress
components are [3]%
\begin{eqnarray}
\rho &=&\frac{b^{\prime }}{r^{2}} \\
p_{r} &=&2\left( 1-\frac{b}{r}\right) \frac{\Phi ^{\prime }}{r^{2}}-\frac{b}{%
r^{3}} \\
p_{\theta } &=&p_{\psi }=\left( 1-\frac{b}{r}\right) \left[ \Phi ^{\prime
\prime }+\Phi ^{\prime 2}+\frac{\Phi ^{\prime }}{r}-\frac{b^{\prime }r-b}{%
2r(r-b)}\Phi ^{\prime }-\frac{b^{\prime }r-b}{2r^{2}(r-b)}\right]
\end{eqnarray}%
where primes denote differentiation with respect to $r$.

By construction, the wormhole geometry has a hole instead of a center and so
we shall change the lower limit of integration in Eq.(3) to the minimum
allowed radius or throat $\ r_{0}$ defined by $b(r_{0})=r_{0}$. The radius $%
r $ has the significance that it is the embedding space radial coordinate;
it decreases from $+\infty $ to $r=r_{0}$ in the lower side and again
increases to $+\infty $ in the upper side. This requires us to change the
integrals (4) and (5) to 
\begin{eqnarray}
Mc^{2} &=&\frac{1}{2}\int_{r_{0}}^{r}T_{0}^{0}r^{2}dr+\frac{r_{0}}{2} \\
E_{M} &=&\frac{1}{2}\int_{r_{0}}^{r}T_{0}^{0}(g_{rr})^{\frac{1}{2}}r^{2}dr
\end{eqnarray}%
where $g_{rr}=\left( 1-\frac{b(r)}{r}\right) ^{-1}$, the entire spacetime
geometry being assumed to be free of singularities. The constant $\frac{r_{0}%
}{2}$ in Eq.(10) comes from the integration of Einstein's equation $\frac{%
\partial M}{\partial r}=\frac{1}{2}T_{0}^{0}r^{2}$ and we shall choose it so
as to offset the inner boundary term $\frac{b(r_{0})}{2}$ coming from the
integration. When $T_{0}^{0}=0$, we should fix $r_{0}=0$ in order to recover 
$M=0$. In geometries with a regular center, one has $r_{0}=0$, the above
then reproduces Eqs.(4) and (5) respectively. The difference between the
above integrals, viz., 
\begin{equation}
\widetilde{E}_{G}=Mc^{2}-E_{M}=\frac{1}{2}\int_{r_{0}}^{r}[1-(g_{rr})^{\frac{%
1}{2}}]T_{0}^{0}r^{2}dr+\frac{r_{0}}{2}
\end{equation}%
is what we call the total gravitational energy of wormholes within the
region of integration. Clearly, it is a straightforward adaptation of Eq.(3)
to wormholes. However, one immediately notices that due to the presence of
the nonzero last term, the sign of $T_{0}^{0}$ does not necessarily
determine the sign of $\widetilde{E}_{G}$, as would be the case otherwise.
Eq.(12) is the main proposition of our paper and will be implemented in the
sequel.

In Ref.[17] it is shown that the gravitational energy density in a
stationary attractive gravity field can be written in remarkable analogy
with electrical energy density of Maxwell electrodynamics: The total
gravitational energy $E_{G}$ can be written as a volume integral of a
perfect square of the gravitational field strength $F_{G}$, that is, $%
E_{G}=-\int_{0}^{\infty }F_{G}^{2}dV$ where $dV$ is the element of spatial
3-volume. In case of wormhole spacetime, the expression for $\widetilde{E}%
_{G}$ can be re-written as (see Appendix): 
\begin{equation}
\widetilde{E}_{G}=-\int_{r_{0}}^{\infty }\frac{1}{r^{2}}\left[ 1-\left(
g_{rr}\right) ^{-\frac{1}{2}}\right] ^{2}dV+r_{0}=-\int_{r_{0}}^{\infty }%
\widetilde{F}_{G}^{2}dV+r_{0}.
\end{equation}%
Furthermore, Lynden-Bell \textit{et al} [17] have shown that one can then
introduce a function $\Psi $ defined by%
\begin{equation}
\widetilde{F}_{G}\equiv \frac{1}{r}\left[ 1-\left( g_{rr}\right) ^{-\frac{1}{%
2}}\right] =\pm \left( g_{rr}\right) ^{-\frac{1}{2}}\frac{\partial \Psi }{%
\partial r}.
\end{equation}%
and that there is then a change of coordinates that will make the spatial
slices conformally flat with conformal factor $e^{2\Psi }$. The $\pm $ sign
in Eq.(14) corresponds respectively to repulsive and attractive nature of
gravitational potential $\Psi $ and the choice is generally open unless the
information is provided by independent physical observations.

A typical wormhole solution may be derived from source ($T_{\nu }^{\mu }$)
that has before it an overall wrong sign (negative) so that all energy
conditions are violated. A well known example is the Ellis-Bronnikov
wormhole. Then the equation for $\Psi $ [\textit{Eq.(9)} of [17], (A6)
below] should be rephrased as%
\begin{equation}
\nabla ^{2}\Psi =-\frac{1}{2}T_{0}^{0}+\frac{1}{2}(\nabla \Psi )^{2}
\end{equation}%
which shows that a positive gravitational energy density $\frac{1}{2}(\nabla
\Psi )^{2}$ is acting alongside negative exotic matter density ($-\frac{1}{2}%
T_{0}^{0}$) as a source of $\Psi $.

\begin{center}
\textbf{3. Conformal factor interpretation}
\end{center}

A good example to demonstrate the conformal factor interpretation is the
Ellis III wormhole which is a solution of Einstein minimally coupled
equation with an overall negative source term. It has the metric [15]

\begin{equation}
ds^{2}=-f(l)dt^{2}+\frac{1}{f(l)}\left[ dl^{2}+(l^{2}+m^{2})\left( d\theta
^{2}+\sin ^{2}\theta d\psi ^{2}\right) \right] ,
\end{equation}

\begin{equation}
f(l)=\exp [-2\beta \{\frac{\pi }{2}-\arctan (\frac{l}{m})\}]
\end{equation}%
\begin{equation}
\varphi (l)=\left[ \sqrt{2}\sqrt{1+\beta ^{2}}\right] \left( \frac{\pi }{2}%
-\arctan (\frac{l}{m})\right)
\end{equation}%
where $m$ and $\beta $ are two constant arbitrary parameters. The throat
appears at $l_{0}=m\beta $. The spacetime (16) is singularity free and
Taylor expansion of $f(l)$ gives asymptotic mass-energy $M^{+}=m\beta $ on
one side and $M^{-}=-m\beta e^{\beta \pi }$ on the other. These masses
follow from the definition (10) as well. For a recent study of geodesics in
the $\beta =0$ case, see [26] and for its stability, see [27]. A remarkable
feature of this solution is that the parameters can be adjusted to make the
wormhole both macroscopic and microscopic satisfying quantum energy
conditions [28,29].

The metric (16) can be rewritten in the standard MTY form by redefining the
radial variable as [30]%
\begin{equation}
r^{2}=(l^{2}+m^{2})\exp [2\beta \{\frac{\pi }{2}-\arctan (\frac{l}{m})\}]
\end{equation}%
(Note that $l\rightarrow \pm \infty $ implies $r\rightarrow +\infty $ and
conversely). Then the redshift function $\Phi (r)$ is given by%
\begin{equation}
\Phi (r)=\beta \left[ \arctan \left\{ \frac{l(r)}{m}\right\} -\frac{\pi }{2}%
\right]
\end{equation}%
and the shape function $b(r)$ is given by 
\begin{equation}
b(r)=r\left[ 1-\frac{[l(r)-m\beta ]^{2}}{r^{2}}\exp [2\beta \{\frac{\pi }{2}%
-\arctan (\frac{l(r)}{m})\}]\right]
\end{equation}%
such that $\frac{b(r)}{r}\rightarrow 0$ as $r\rightarrow \infty $. Throat
occurs at the minimum of $r$ where $b(r_{0})=r_{0}$. Putting $l_{0}=m\beta $
in Eq.(19), we find 
\begin{equation}
r_{0}=m(1+\beta ^{2})^{\frac{1}{2}}\exp [\beta \{\frac{\pi }{2}-\arctan
\beta \}].
\end{equation}%
Now consider the following transformation $l\rightarrow R$ where $R$ is the
isotropic radial coordinate 
\begin{equation}
l=\frac{R^{2}-m^{2}}{2R}
\end{equation}%
with its inverse%
\begin{equation}
R=l+\sqrt{l^{2}+m^{2}}.
\end{equation}%
It can be verified that the original metric (16) goes into its isotropic
form as follows%
\begin{eqnarray}
ds^{2} &=&-f[l(R)]dt^{2}+\frac{1}{f}\left[ \frac{R^{2}+m^{2}}{2R^{2}}\right]
^{2}\left[ dR^{2}+R^{2}\left( d\theta ^{2}+\sin ^{2}\theta d\psi ^{2}\right) %
\right] \\
&=&-f[l(R)]dt^{2}+e^{2\Psi (R)}dl_{E}^{2}
\end{eqnarray}%
giving the conformal factor 
\begin{equation}
\Psi (R)=\frac{1}{2}\ln \left\{ \frac{1}{f}[l(R)]\left[ \frac{R^{2}+m^{2}}{%
2R^{2}}\right] ^{2}\right\} .
\end{equation}%
Putting this $\Psi $ in Eq.(14) and integrating as in Eq.(13) plus a bit of
algebra taking care of coordinate changes $r\rightarrow l\rightarrow R$, we
get the same expression for $\widetilde{E}_{G}$ as defined in Eq.(12) with $%
T_{0}^{0}=\rho $ from Eq.(28) below. By itself, the result is no surprise as
the calculation in Ref.[17] is quite generic. What is of interest here is
that the Maxwell analogy is valid even in the case of exotic matter
associated with geometry peculiar to wormholes. This exercise lends
reliability to $\widetilde{E}_{G}$ as defined in Eq.(12). We now exemplify
that the definition is quite robust as well.

\begin{center}
\textbf{4. Robustness of }$\widetilde{E}_{G}$
\end{center}

To show it, we allow deviations from the conditions behind the original
definition of $E_{G}$ in Eq.(3), the wormhole reincarnation of which is $%
\widetilde{E}_{G}$. The same wormhole metric (16) is a good candidate for
this purpose. Our aim is to see if $\widetilde{E}_{G}$ still produces known
results. The deviation lies in the following features. In (16), the
matter-energy content is not localized in a finite region though the stress
quantities do fall off to zero with radial distance. As required of
wormholes, the solution has two asymptotically flat regions ($+ve$: $l\in
\lbrack m\beta ,+\infty )$) and ($-ve$: $l\in \lbrack m\beta ,-\infty )$)
connected by the throat at $l_{0}=m\beta $. The stress components, given
below, identically vanish in the asymptotic limit $l\rightarrow \pm \infty $%
. Thus, using Eqs.(20) and (21) in Eqs.(7)-(9), we obtain%
\begin{eqnarray}
\rho &=&-\frac{m^{2}(1+\beta ^{2})}{(l^{2}+m^{2})^{2}}e^{-\beta \lbrack \pi
-2\arctan (\frac{l}{m})]} \\
p_{r} &=&\rho \\
p_{\theta } &=&p_{\psi }=-\rho
\end{eqnarray}%
which shows that both WEC\ \ and NEC violated everywhere since $\rho <0$ and 
$\rho +p_{r}=2\rho <0$ respectively. Using Eqs.(28), (29), (21), (22) and
definitions (10) and (11), we get on the $+ve$ side, noting that $\rho
r^{2}dr=db$, $b(+\infty )=m\beta $, $b(r_{0})=r_{0}=m\beta $: 
\begin{eqnarray}
M^{+}c^{2} &=&\frac{1}{2}\int_{r_{0}}^{\infty }T_{0}^{0}r^{2}dr+\frac{r_{0}}{%
2}=m\beta \\
E_{M}^{+} &=&\frac{1}{2}\int_{r_{0}}^{\infty }T_{0}^{0}(g_{rr})^{\frac{1}{2}%
}r^{2}\frac{dr}{dl}dl \\
&=&\frac{m}{2}\left( \frac{1+\beta ^{2}}{\beta }\right) \left( 1-\sqrt{%
e^{\beta (\pi -2\arctan (\beta )}}\right) .
\end{eqnarray}%
We always find that $\widetilde{E}_{G}^{+}=M^{+}c^{2}-E_{M}^{+}>0$ or
repulsive gravity for $m>0$, $\beta >0$. Proceeding in similar manner for
the other side, we get%
\begin{eqnarray}
M^{-}c^{2} &=&-m\beta e^{\beta \pi } \\
E_{M}^{-} &=&\frac{m}{2}\left( \frac{1+\beta ^{2}}{\beta }\right) \left(
e^{\beta \pi }-\sqrt{e^{\beta (\pi -2\arctan (\beta )}}\right)
\end{eqnarray}%
showing that $\widetilde{E}_{G}^{-}=M^{-}c^{2}-E_{M}^{-}<0$ or attractive
gravity for $m>0$, $\beta >0$. We recall that the Ellis solution (16)
describes a Janus-faced wormhole that sucks in test particles in one mouth
and pumps out at the other. The $\widetilde{E}_{G}^{\pm }$ just calculated
nicely describe this scenario despite the deviations mentioned above.

We can have some additional insight about the wormhole with zero Keplerian
mass, $M=m\beta =0\Rightarrow \beta =0$, for which the metric can be written
in standard coordinates as 
\begin{equation}
ds^{2}=-dt^{2}+\frac{dr^{2}}{1-\frac{m^{2}}{r^{2}}}+r^{2}(d\theta ^{2}+\sin
^{2}\theta d\psi ^{2})
\end{equation}%
where $r^{2}=l^{2}+m^{2}$. The shape function is $b(r)=\frac{m^{2}}{r}$ and
the throat appears at $r_{0}=m$ or equivalently at $l_{0}=0$. This is a well
discussed single parameter symmetric wormhole made entirely of the massless
scalar field $\varphi $ [cf. Eqs.(17),(18)]. We obtain from above, in the
limit $\beta \rightarrow 0$, 
\begin{eqnarray}
M^{+}c^{2} &=&0,E_{M}^{+}=-\frac{m\pi }{4}\Rightarrow \widetilde{E}_{G}^{+}=%
\frac{m\pi }{4} \\
M^{-}c^{2} &=&0,E_{M}^{-}=\frac{m\pi }{4}\Rightarrow \widetilde{E}_{G}^{-}=-%
\frac{m\pi }{4}.
\end{eqnarray}%
We see that $\widetilde{E}_{G}^{+}>0$ and $\widetilde{E}_{G}^{-}<0$, and
vanishing mass-energy $M^{\pm }c^{2}$ on both sides, but nonvanishing $%
\widetilde{E}_{G}^{\pm }$ contributed by the scalar field $\varphi $. The
nonvanishing of $\widetilde{E}_{G}^{\pm }$ explains why the wormhole is able
to capture test particles despite the fact that it has zero Keplerian mass
[15].

\begin{center}
\textbf{5. Energetics in the Mazur-Mottola star}
\end{center}

Consider the static spherically symmetric vacuum condensate star (also
called gravastar) devised by Mazur and Mottola [23]. The star has an
isotropic de-Sitter vacuum in the interior, the matter \textit{marginally }%
satisfying the NEC and strictly violating the Strong Energy Condition (SEC) $%
\rho +3p\geq 0$. The star has an interior boundary at $r=r_{1}$ containing
de Sitter vacuum ($p=-\rho $) and an exterior boundary at $r=r_{2}$ beyond
which the spacetime is described by the Schwarzschild exterior ($p=0$, $\rho
=0$) of mass $M$. The intermediate region is covered by a thin shell of
stiff matter ($p=+\rho $).

The self-consistent interior de Sitter metric for a constant density vacuum $%
\rho =\rho _{vac}=3H_{0}^{2}/8\pi G=const.>0$ is given by%
\begin{equation}
d\tau ^{2}=-\left( 1-\frac{r^{2}}{\widehat{R}^{2}}\right) dt^{2}+\left( 1-%
\frac{r^{2}}{\widehat{R}^{2}}\right) ^{-1}dr^{2}+r^{2}\left( d\theta
^{2}+\sin ^{2}\theta d\psi ^{2}\right)
\end{equation}%
where $\widehat{R}^{2}=\frac{3}{8\pi G\rho _{vac}}=\frac{1}{H_{0}^{2}}$. The
transverse pressures in the thin shell serve to act more like a Roman arch
supporting the star than making any substantial contribution to mass-energy.
The shell contribution has actually been shown [23] to be negligible, $%
M_{shell}\sim \epsilon M$ where $0<\epsilon \ll 1$. \ Israel-Darmois
junction conditions then imply a negative surface tension at the inner
interface of the shell which balances the outward force exerted by the
repulsive vacuum within. Likewise, the positive surface tension at the outer
interface balances the inward force from without. Using the thin shell
approach, Visser and Wiltshire [31] studied dynamic stability of similar
type of configurations. The mass-energy contained within the boundary radius 
$r=r_{b}$ is given by 
\begin{equation}
M=\frac{4\pi }{3}r_{b}^{3}\rho _{vac}>0.
\end{equation}%
Physics begins to become interesting in the region where horizon $r_{hor}$
ought to have formed. This is the region where the inner and outer
boundaries tend to meet, viz.,%
\begin{equation}
r_{1}\sim r_{2}\sim 2M\sim \frac{1}{H_{0}}=\widehat{R}\sim r_{hor}
\end{equation}%
at which the $g_{rr}$ from either side tend to approach arbitrarily close to
infinity. Since $r_{0}=0$ (the star has a regular center), our $\widetilde{E}%
_{G}$ coincides with $E_{G}$. Thus, putting $g_{rr}=\left( 1-\frac{r^{2}}{%
\widehat{R}^{2}}\right) ^{-1}$ and $T_{0}^{0}=\rho _{vac}$ in Eq.(12), we
get the exact expression: 
\begin{eqnarray}
E_{G} &=&\frac{1}{2}\int_{0}^{r_{b}}[1-(g_{rr})^{\frac{1}{2}%
}]T_{0}^{0}r^{2}dr  \nonumber \\
&=&\rho _{vac}\left[ \frac{r_{b}^{3}}{6}-\frac{\widehat{R}}{4}\left\{ 
\widehat{R}^{2}\arcsin \left( \frac{r_{b}}{\widehat{R}}\right) -r_{b}\sqrt{%
\widehat{R}^{2}-r_{b}^{2}}\right\} \right] .
\end{eqnarray}%
Taking the boundary close to the horizon, viz., $r_{b}\rightarrow $ $%
\widehat{R}\sim r_{hor}$, we find that (in units $8\pi G=1$): 
\begin{eqnarray}
E_{G} &=&\left( \frac{4-3\pi }{24}\right) \rho _{vac}\widehat{R}^{3} \\
&=&\left( \frac{4-3\pi }{24}\right) \times 3\widehat{R}=-0.678\left( \frac{1%
}{H_{0}}\right) =-1.356M.  \nonumber
\end{eqnarray}%
The result $E_{G}<0$ implies that the total gravitational energy inside the
de Sitter star is attractive whereas independent physical information is
that the de Sitter space has repulsive gravity (because $\rho +3p<0$). So
one might conclude that the sign of $E_{G}$ is conveying a wrong result.
This need not be so. We have to recall that the de Sitter expansion means
that the entire 3-space is expanding. On the other hand, by construction the
de Sitter gravastar has a finite boundary close to the horizon of an
exterior Schwarzschild metric, the inner boundary exerting inward force
balancing the outward force from within.

The whole scenario can be given a \textit{metric} \textit{equivalent}
description replacing the inner region by the interior Schwarzschild
solution for constant density $\rho _{vac}$. That this replacement is indeed
possible can be seen by looking at the interior Schwarzschild $g_{rr}$
(Remember: for $E_{G}$ we need to consider only $g_{rr}$) which is given by%
\begin{equation}
g_{rr}=\left( 1-\frac{2Mr^{2}}{r_{b}^{3}}\right) ^{-1}
\end{equation}%
which matches the exterior at $r=r_{b}$. Putting $M$ from Eq.(40) and using $%
\widehat{R}^{2}=\frac{3}{8\pi G\rho _{vac}}$, we get exactly the $g_{rr}$ of
metric (39). The interior Schwarzschild metric always has $E_{G}<0$. By the
same token, the gravastar too can have attractive gravity in the interior
via the interpretation of metric equivalence. This explanation seems
feasible since the gravastar is after all a stable Schwarzschild-like star
(as viewed from outside) with an \textit{arbitrarily }thin layer of
quasi-normal matter at a place where horizon would have formed. In the next
sections, we shall consider truncated wormholes which are constructed in a
manner very similar to that of gravastar.

\begin{center}
\textbf{6. Thin shell contribution}
\end{center}

Several asymptotically flat wormholes are known in the literature with
matter threading the wormhole all the way to infinity with radial fall-offs
in the stress quantities. Such wormholes might be existing in nature as an
end result of some past astrophysical phenomena or might be artificially
constructed by truncation. For completeness, we shall calculate the thin
shell contribution to $\widetilde{E}_{G}$ although the contribution can be
made arbitrarily small.

The idea of a truncated wormhole is the following. One wants to artificially
create a wormhole by localizing the exotic matter within a finite radius
around the throat $r=r_{0}$ of a given solution. This can be achieved by
taking a cut-off at any finite radius away from $r=r_{0}$, say, at $%
r=a>r_{0} $ and matching the surface at $r=a$ to an exterior Schwarzschild
vacuum. The matching brings into play junction conditions as follows:\ The
induced metric on the spacelike junction interface $\Sigma $ is given by%
\begin{equation}
ds_{\Sigma }^{2}=-d\tau ^{2}+a^{2}\left( d\theta ^{2}+\sin ^{2}\theta d\psi
^{2}\right)
\end{equation}%
where $\tau $ is the proper time on the surface. On this surface the matter
energy density $\sigma $ and transverse pressures \ are calculated from the
jump in the extrinsic curvature $[K_{ij}]_{-}^{+}=K_{ij}^{+}-K_{ij}^{-}$ as $%
r\rightarrow a\pm $. The result is [32]%
\begin{eqnarray}
\sigma &=&-\frac{1}{4\pi a}\left[ \sqrt{1-\frac{2M}{a}}-\sqrt{1-\frac{b(a)}{a%
}}\right] \\
P_{\theta } &=&P_{\psi }=\frac{1}{8\pi a}\left[ \frac{1-\frac{M}{a}}{\sqrt{1-%
\frac{2M}{a}}}-\zeta \sqrt{1-\frac{b(a)}{a}}\right]
\end{eqnarray}%
where $\zeta =1+a\frac{d\Phi }{dr}\mid _{r=a}$, $\sigma $ is the surface
energy density and $P_{\theta }$, $P_{\psi }$ are transverse pressures on
the surface. When $\sigma =0$, $P_{\theta }=P_{\psi }=0$, the surface $r=a$
is called the boundary. However, a more interesting possibility is to
consider an arbitrarily\ thin shell of quasi-normal matter (that is, matter
satisfying both WEC and NEC) at $r=a$. Then the total mass-energy is given
by [8] 
\begin{equation}
M=\frac{b(a)}{2}+M_{shell}\left[ \sqrt{1-\frac{b(a)}{a}}-\frac{M_{shell}}{2a}%
\right]
\end{equation}%
where $M_{shell}=4\pi a^{2}\sigma $ is the shell mass contribution. If $%
b(a)=2M$, then $\sigma =0$. To have an idea of how $\sigma \neq 0$
contributes to the gravitational energy $\widetilde{E}_{G}$, we should fix
the shape function $b(a)$ to a value slightly away from $2M$. For instance,
we can fix $b(a)=2M-\epsilon M$ where $0<\epsilon \ll 1$ is a dimensionless
parameter related to the infinitesimally thin thickness of the shell. In
this case we get, to leading order in $\epsilon $,%
\begin{equation}
M_{shell}\simeq \frac{\epsilon M}{2}.
\end{equation}%
Up to a factor ($\frac{1}{2}$), this is exactly the same result as that
obtained in Ref.[23]. To get an idea of the measure of $E_{M}$ in the shell,
we can regard the density to be approximately the constant $\sigma $
throughout the shell while the spacetime can be approximately described by a
Schwarzschild metric for mass $M.$ Then 
\begin{eqnarray}
E_{M}^{shell} &=&\frac{1}{2}\sigma \int_{a}^{a+\epsilon }\left( 1-\frac{2M}{r%
}\right) ^{-\frac{1}{2}}r^{2}dr \\
&\simeq &\frac{\epsilon M}{2}.\frac{\epsilon }{4\pi a^{2}}\left[ 3a^{2}+aM%
\right] =O(\epsilon ^{2})
\end{eqnarray}%
The total gravitational energy of the truncated wormhole therefore becomes%
\begin{eqnarray}
\widetilde{E}_{G} &=&M-E_{M}=\frac{1}{2}\int_{r_{0}}^{a}[1-(g_{rr})^{\frac{1%
}{2}}]T_{0}^{0}r^{2}dr+\frac{r_{0}}{2} \\
&&+M_{shell}\left[ \sqrt{1-\frac{b(a)}{a}}-\frac{M_{shell}}{2a}\right]
-E_{M}^{shell}.  \nonumber
\end{eqnarray}%
The term $M_{shell}^{2}$ as well as $E_{M}^{shell}$ may be neglected as
being of order $\epsilon ^{2}$. To calculate $M_{shell}$ for a given shape
function $b(r)$, we express Eq.(49) in terms of $b(a)$ as follows%
\begin{equation}
M_{shell}=\frac{\epsilon }{2}\left[ \frac{b(a)}{2-\epsilon }\right] \simeq 
\frac{\epsilon b(a)}{4}
\end{equation}%
to first order in $\epsilon $. So the contribution to mass-energy coming
from the thin shell reduces to $\frac{\epsilon b(a)}{4}$ which is always
positive. This may be added to the right hand side of Eq.(12). So, in all,
we can write 
\begin{equation}
\widetilde{E}_{G}=M-E_{M}=\frac{1}{2}\int_{r_{0}}^{a}[1-(g_{rr})^{\frac{1}{2}%
}]T_{0}^{0}r^{2}dr+\frac{r_{0}}{2}+\frac{\epsilon b(a)}{4}.
\end{equation}%
The contribution to energy from thin shell of quasi-normal matter is
essentially of academic interest rather than anything substantial because of
the limit $\epsilon \rightarrow 0$ and is generally ignored. (See for
instance the second reference in [32].) We too shall ignore it in what
follows.

The physical situation in any wormhole is that the cross-sectional area of a
bundle of light rays entering one mouth must decrease and then increase
while emerging at the other mouth. This can be produced only by the
gravitational repulsion of matter [20] at or in the vicinity of the throat.
Let us now analyze a couple of known truncated wormhole solutions to see if
this criterion is satisfied by the definition of $\widetilde{E}_{G}$.

\begin{center}
\textbf{7. Lobo phantom wormhole }
\end{center}

The metric is given by 
\begin{equation}
ds^{2}=-\left[ 1-\left( \frac{r_{0}}{r}\right) ^{1-\alpha }\right] ^{\frac{%
1+\alpha \omega }{1-\alpha }}dt^{2}+\frac{dr^{2}}{1-\left( \frac{r_{0}}{r}%
\right) ^{1-\alpha }}+r^{2}\left( d\theta ^{2}+\sin ^{2}\theta d\psi
^{2}\right)
\end{equation}%
where $\alpha $ and $\omega $ are constant parameters and $0<\alpha <1$. The
phantom equation of state further demands that $p_{r}/\rho =\omega <-1$. The
shape function and the redshift function respectively are%
\begin{eqnarray}
b(r) &=&r^{\alpha }r_{0}^{1-\alpha } \\
\Phi (r) &=&\left( \frac{1+\alpha \omega }{1-\alpha }\right) \ln \left[
1-\left( \frac{r_{0}}{r}\right) ^{1-\alpha }\right] .
\end{eqnarray}%
The density and radial pressure for this wormhole are%
\begin{eqnarray}
\rho &=&\frac{\alpha r_{0}}{r^{3}}\left( \frac{r_{0}}{r}\right) ^{-\alpha }
\\
p_{r} &=&\frac{\alpha \omega r_{0}}{r^{3}}\left( \frac{r_{0}}{r}\right)
^{-\alpha }.
\end{eqnarray}%
To have the spacetime free of singularities, we must impose a constraint $%
1+\alpha \omega =0$. We thus obtain the NEC violating condition%
\begin{equation}
\rho +p_{r}=\frac{\alpha r_{0}}{r^{3}}\left( \frac{r_{0}}{r}\right)
^{-\alpha }(\frac{\alpha -1}{\alpha })<0
\end{equation}%
satisfied for all $r$.

As discussed by Lobo [8], this wormhole can be truncated at some finite
radius at $r=a$ away from the throat $r=r_{0}$ to match to an exterior
Schwarzschild spacetime. Neglecting the thin shell contribution $O(\epsilon )
$, we can explicitly do the integration in Eq.(12) to get $\widetilde{E}_{G}$%
. Taking for example, $\alpha =\frac{1}{3}$, so that $\omega =-3$, we obtain%
\begin{equation}
\rho =\frac{1}{3}\left( \frac{1}{r}\right) ^{-\frac{1}{3}}\frac{1}{r^{3}}>0.
\end{equation}%
Choosing mass units in which $r_{0}=1$ and using the metric (55), we get 
\begin{eqnarray}
\widetilde{E}_{G} &=&\frac{1}{2}\int_{1}^{a}\left[ 1-\left( 1-r^{-\frac{2}{3}%
}\right) ^{-\frac{1}{2}}\right] \rho r^{2}dr+\frac{1}{2} \\
&=&\frac{a^{\frac{1}{3}}}{2}\left[ a^{-\frac{2}{3}}-\sqrt{1-a^{-\frac{2}{3}}}%
\right] .
\end{eqnarray}%
For any value of $a\geq 1$, it is evident that $\widetilde{E}_{G}>0$
(Fig.1). That is, there is the expected repulsion around the throat, and
elsewhere within the cut off boundary. One may take any other value in the
range $0<\alpha <1$ and $\omega <-1$ consistent with $1+\alpha \omega =0$ to
see that the same repulsion continues to occur.

\begin{center}
\textbf{8. Lemos - Lobo - Oliveira wormhole (LLO) }
\end{center}

The metric inside $r_{0}\leq r\leq a$ is given by [33] 
\begin{equation}
ds^{2}=-\left[ 1-\left( \frac{r_{0}}{a}\right) ^{\frac{1}{2}}\right] dt^{2}+%
\frac{dr^{2}}{1-\left( \frac{r_{0}}{r}\right) ^{\frac{1}{2}}}+r^{2}\left(
d\theta ^{2}+\sin ^{2}\theta d\psi ^{2}\right) ,
\end{equation}%
which gives 
\begin{eqnarray}
b(r) &=&\sqrt{rr_{0}} \\
\Phi &=&\frac{1}{2}\ln \left[ 1-\left( \frac{r_{0}}{a}\right) ^{\frac{1}{2}}%
\right] =const.
\end{eqnarray}%
The exterior vacuum is described by the Schwarzschild metric in $a\leq
r<\infty $ as follows%
\begin{equation}
ds^{2}=-\left[ 1-\frac{(r_{0}a)^{\frac{1}{2}}}{r}\right] dt^{2}+\frac{dr^{2}%
}{1-\frac{(r_{0}a)^{\frac{1}{2}}}{r}}+r^{2}\left( d\theta ^{2}+\sin
^{2}\theta d\psi ^{2}\right) .
\end{equation}%
The energy density and radial pressure are 
\begin{eqnarray}
\rho &=&\frac{1}{r^{2}}\frac{db}{dr}=\frac{\sqrt{r_{0}}}{2r^{5/2}}>0, \\
p_{r} &=&-\frac{\sqrt{r_{0}}}{r^{5/2}}<0 \\
\frac{p_{r}}{\rho } &=&-2.
\end{eqnarray}%
We have a phantom equation of state ($\omega =-2$) here although the metric
properties are quite different from the earlier example. NEC is violated
everywhere, including at the throat, since $\rho +p_{r}=-$ $\frac{\sqrt{r_{0}%
}}{2r^{5/2}}$. The throat appears at $r=r_{0}$, and the spacetime is
perfectly regular there. To get an estimate, we again choose mass units in
which $r_{0}=1$ with the cut-off at $r=a$. Using the metric (64), we can
calculate $\widetilde{E}_{G}$ as follows: 
\[
\widetilde{E}_{G}=\frac{1}{2}\int_{1}^{a}\left[ 1-\left( 1-r^{-\frac{1}{2}%
}\right) ^{-\frac{1}{2}}\right] \rho r^{2}dr+\frac{1}{2} 
\]%
\[
=\left( \frac{1}{2a^{\frac{1}{4}}\sqrt{1-a^{-\frac{1}{2}}}}\right) [a^{\frac{%
1}{4}}+a^{\frac{3}{4}}\left( \sqrt{1-a^{-\frac{1}{2}}}-1\right) 
\]%
\begin{equation}
-\left( \sqrt{a^{\frac{1}{2}}-1}\right) \ln \left( a^{\frac{1}{4}}+\sqrt{a^{%
\frac{1}{2}}-1}\right) ].
\end{equation}%
From Fig.1, it is evident that $\widetilde{E}_{G}>0$ for $1<a<2.15$ while $%
\widetilde{E}_{G}\leq 0$ for $a\geq 2.15$. One also sees exactly where $%
\widetilde{E}_{G}$ changes sign. The main thing however is that there is the
desired repulsion (defocussing) in the vicinity of the throat which lie
within the range $1<a<2.15$.

\begin{center}
\textbf{9. Summary}
\end{center}

The original derivation of the formula for $E_{G}$ for a static spherically
symmetric asymptotically flat spacetime, as given in [17,18], is adapted to
exotic matter sources that automatically satisfy local conservation laws.
There is a statement in [19] to the effect that $E_{G}<0$ for localized
sources satisfying energy conditions. The statement is certainly true for
ordinary fluids. However, the converse question, namely, whether $E_{G}>0$
in case of energy condition violating matter such as occurring in gravastar
or wormholes, remained essentially open. The present article is a primary
initiative to answer the question.

To handle wormhole configurations, which require repulsion, we proposed the
expression $\widetilde{E}_{G}$ consistent with wormhole geometry without
center. Subsequent implementation of it not only supported the Maxwellian
analogy in a wider regime but also correctly produced the gravitational
energy picture in the Ellis, Lobo and LLO phantom wormholes. The definition
of $\widetilde{E}_{G}$ was also shown to be robust in the sense that it did
reproduce the expected behavior under slightly deviating circumstances. The
explicit analysis of truncated wormhole lends force to the notion that a
condition \textit{weaker} than WEC violation, namely, NEC\ violation is
sufficient to cause defocussing of light rays. The Mazur-Mottola gravastar
does not have wormhole topology but the exact result for $\widetilde{E}_{G}$
($\equiv E_{G}$) admits a plausible physical interpretation.

What are the possible implications of these results? We recall that $\rho >0$
wormholes are not ruled out [3] but there needs to be defocussing of light
rays, hence repulsion, at or in the vicinity of the throat. Looking at
Eq.(12) we realize that the integral can, in principle, result in values
having either signs depending on the wormhole model chosen. If it so happens
that the integral is large and negative overcoming the additive factor $%
\frac{r_{0}}{2}$, then we end up with $\widetilde{E}_{G}<0$ or lack of
repulsion everywhere. We might rule out such wormhole configurations as
physically unrealistic or unrealizable, though they might be technically
valid solutions.

(Note added: It has been brought to our notice that wormholes in ghost
scalar field theories are unstable under both linear and nonlinear
perturbations [34-36], which refutes the result of Ref. [27]).

\textbf{Acknowledgments}

We wish to thank Guzel N. Kutdusova of the Science Research Administration
of BSPU where part of the work was carried out. Thanks are due to Benjamin
I. Lye, Wang Hai Ni and Aydar R. Bikmetov for helpful comments. KKN thanks
ITP, CAS for warm hospitality and TWAS, Italy for financial support. We
thank two anonymous referees for their insightful comments.

\bigskip

\textbf{References}

[1] A. Einstein and N. Rosen, Phys. Rev. \textbf{48}, 73 (1935).

[2] L. Flamm, Z. Physik \textbf{17}, 448 (1916).

[3] M.S. Morris and K.S. Thorne, Am. J. Phys. \textbf{56}, 395 (1988); M.S.
Morris, K.S. Thorne and U. Yurtsever, Phys. Rev. Lett. \textbf{61}, 1446
(1988).

[4] A.G. Agnese and M. La Camera, Phys. Rev. D \textbf{51,} 2011 (1995);
K.K. Nandi, A. Islam and J. Evans, Phys. Rev. D \textbf{55}, 2497 (1997);
L.A. Anchordoqui, S. P. Bergliaffa, D.F. Torres, Phys. Rev. D \textbf{55}
526 (1997); K.K. Nandi, B. Bhattacharjee, S.M.K. Alam and J. Evans, Phys.
Rev. D \textbf{57}, 823 (1998).

[5] K.A. Bronnikov and J.C. Fabris, Phys. Rev. Lett. \textbf{96}, 251101
(2006) and references therein.

[6] M. Cadoni and M. Cavaglia, Phys. Rev. D \textbf{50}, 6435 (1994); K.K.
Nandi and S.M.K. Alam, Gen. Rel. Grav. \textbf{30}, 1331 (1998); D.N.
Vollick, Class. Quant. Grav. \textbf{16}, 1599 (1999); K.K. Nandi and Y.Z.
Zhang, \ Phys. Rev. D \textbf{70}, 044040 (2004).

[7] K.A. Bronnikov, Grav. Cosmol. \textbf{4}, 49 (1998); C. Barcel\'{o} and
M. Visser, Nucl. Phys. B \textbf{584}, 415 (2000); K.A. Bronnikov and S.-W.
Kim, Phys. Rev. D \textbf{67}, 064027 (2003); E. Rodrigo, Phys. Rev. D 
\textbf{74}, 104025 (2006).

[8] F.S. N. Lobo, Phys. Rev. D \textbf{71}, 084011 (2005); See also: S. V.
Sushkov, Phys.Rev. D \textbf{71}, 043520 (2005); O.B. Zaslavskii, Phys. Rev.
D \textbf{72}, 061303 (2005).

[9] F.S.N. Lobo, Phys. Rev. D \textbf{73}, 064028 (2006).

[10] F. Rahaman, M. Kalam and S. Chakraborty, Gen. Rel. Grav. \textbf{38},
1687 (2006); M. Richarte and C. Simeone, Phys. Rev. D \textbf{76}, 087502
(2007); E.F. Eiroa and C. Simeone, Phys. Rev. D \textbf{76}, 024021 (2007).

[11] D. Hochberg and T.W. Kephart, Phys. Rev. Lett. \textbf{70}, 2665
(1993); C. Barcel\'{o}, Int. J. Mod. Phys. D \textbf{8}, 325 (1999); S.V.
Sushkov and Y.Z. Zhang, Phys. Rev. D \textbf{77}, 024042 (2008).

[12] J.G. Cramer, R.L. Forward, M.S. Morris, M. Visser, G. Benford and G.A.
Landis, Phys. Rev. D \textbf{51}, 3117 (1995); K.S. Virbhadra and G.F.R.
Ellis, Phys. Rev. D \textbf{62}, 084003 (2002); V. Bozza, Phys. Rev. D 
\textbf{66}, 103001 (2002); N.S. Kardashev, I.D. Novikov and A.A. Shatskiy,
Int. J. Mod. Phys. D \textbf{16}, 909 (2007).

[13] K.K. Nandi, Y.Z. Zhang and A.V. Zakharov, Phys. Rev. D \textbf{74},
024020 (2006).

[14] I.Z. Fisher, Zh. Eksp. Teor. Fiz. \textbf{18}, 636 (1948)
[gr-qc/991108]. The solution has been pointed out by Bronnikov and
rediscovered in various forms by several authors afterwards.

[15] H.G. Ellis, J. Math. Phys. 14, \textbf{104 }(1973); \textbf{15}, 520E
(1974).

[16] K.A. Bronnikov, Acta Phys. Polon. B\textbf{\ 4}, 251 (1973).

[17] D. Lynden-Bell, J. Katz and J. Bi\v{c}\'{a}k, Phys. Rev. D \textbf{75},
024040 (2007); Erratum, \textit{ibid}, D \textbf{75}, 044901 (2007).

[18] J. Katz, D. Lynden-Bell and J. Bi\v{c}\'{a}k, Class. Quant. Grav. 
\textbf{23}, 9111 (2006).

[19] J. Katz, Class. Quant. Grav. \textbf{22}, 5169 (2005).

[20] D.N. Page, his comments quoted by MT, the first article of Ref.[3].

[21] J.L. Friedman, K. Schleich and D.M. Witt, Phys. Rev. Lett. \textbf{71},
1486 (1993).

[22] D. Hochberg and M. Visser, Phys. Rev. Lett. \textbf{81}, 746 (1998).

[23] P.O. Mazur and E. Mottola, Proc. Natl. Acad. Sc. USA, \textbf{101},
9545 (2004).

[24] R.M. Wald, \ \textit{General Relativity}, University of Chicago Press,
Chicago (1984).

[25] C.W. Misner, K.S. Thorne and J.A. Wheeler, \textit{Gravitation},
Freeman, San Francisco (1973), pp 467, 603.

[26] T. M\"{u}ller, Phys. Rev. D\textbf{\ 77}, 044043 (2008).

[27] C. Armend\'{a}riz-P\'{\i}con, Phys. Rev. D\textbf{\ 65}, 104010 (2002).

[28] L.H. Ford and T.A. Roman, Phys. Rev. D \textbf{53}, 5496 (1996) and
references therein.

[29] K.K. Nandi, Y.Z. Zhang and K.B. Vijaya Kumar, Phys. Rev. D \textbf{70},
064018 (2004).

[30] K.K. Nandi, I. Nigmatzyanov, R. Izmailov and N.G. Migranov, Class.
Quant. Grav. \textbf{25}, 165020 (2008).

[31] M. Visser and D.L. Wiltshire, Class. Quant. Grav. \textbf{21}, 1135
(2004).

[32] M. Visser, \textit{Lorentzian Wormholes-From Einstein to Hawking, }AIP
Press, NY (1995); M. Visser, S. Kar, N. Dadhich, Phys. Rev. Lett. \textbf{90}%
, 201102 (2003). See also: O. Zaslavskii, Phys. Rev. D \textbf{75}, 084030
(2007).

[33] J.P.S. Lemos, F.S.N. Lobo and S.Q. de Oliveira, Phys. Rev. D \textbf{68}
064004 (2003).

[34] Hisa-aki Shinkai, Sean A. Hayward, Phys. Rev. D \textbf{66}, 044005
(2002).

[35] J.A. Gonzalez, \ F.S. Guzman, O. Sarbach, [arXiv:0806.1370] and
[arXiv:0806.0608], both to appear in Class. Quant. Grav.

[36] K.A. Bronnikov and A.A. Starobinsky, JETP\ Lett, \textbf{85}, 1 (2007).

\begin{center}
\textbf{Appendix }
\end{center}

The basic idea of Lynden-Bell \textit{et al} [17] is to draw an energy
analogy between electrodynamics and general relativity: The total electrical
energy $E_{em}$ of a spherical charge distribution $Q(r)$ can be derived in
various ways but the true electrical energy density $F_{em}^{2}$ ($\equiv 
\frac{Q^{2}}{r^{4}}$) can be found only from the expression which due to
Maxwell, and given by%
\[
E_{em}=\int F_{em}^{2}dV=\int \left( \frac{Q}{r^{2}}\right) ^{2}dV 
\]%
where $dV$ is the elementary volume of flat 3-space. The integral over the
perfect square evidently gives the electrical field strength $F_{em}=\frac{Q%
}{r^{2}}$.

The question is whether a similar notion of gravitational energy density can
be developed within the framework of general relativity. Misner, Thorne and
Wheeler [25] deny the existence of localized gravitational field energy
density in general. Nevertheless they give an expression for it but only in
the exceptional case of spherical symmetry. Lynden-Bell \textit{et al }%
developed the gravitational field energy density in a form which is
remarkably analogous to the above Maxwell expression. They further extended
the notion to axisymmetric spacetimes.

Adapting their derivation to spherically symmetric wormholes, we define $x=%
\frac{2M(r)}{r}=\frac{b(r)}{r}$, and noting $x\rightarrow 0$ as $%
r\rightarrow \infty $, and $x=1$ at the throat $r=r_{0}$, and further using $%
dV=(1-x)^{-\frac{1}{2}}\times 4\pi r^{2}dr$, we obtain 
\[
\widetilde{E}_{G}=-\int_{r_{0}}^{\infty }\widetilde{F}_{G}^{2}dV+r_{0} 
\]%
which is Eq.(13) in the text.

\end{document}